\documentclass[12pt]{article}

\usepackage{newtxtext,newtxmath}
\newcommand{\RSLN}{MoLiN}
\usepackage{graphicx}

\usepackage{todonotes}

\usepackage[letterpaper,margin=1in]{geometry}

\renewenvironment{abstract}
	{\quotation}
	{\endquotation}

\date{}

\makeatletter
\renewcommand{\fnum@figure}{\textbf{Figure \thefigure}}
\renewcommand{\fnum@table}{\textbf{Table \thetable}}
\makeatother

\usepackage{scicite}

\usepackage{url}

\def\scititle{
    Linking extended vector wave fields with momentum space topology
}
\title{\bfseries \boldmath \scititle}

\author{
	A.~Neuhaus$^{1\dagger}$,
	P.~Gessler$^{1\dagger}$,
    P.~Dreher$^{1,2\dagger}$,
    D. Janoschka$^1$, 
    A. R\"odl$^1$, 
    M. Manten$^1$, \and
    Th. Bauer$^{3,4}$, 
    M. Azhar$^{1}$,
    B. Frank$^5$, 
    T.J. Davis$^{1,5,6}$, 
    H. Giessen$^5$, \and
    K. Everschor-Sitte$^{1\ast}$, 
    F. Meyer zu Heringdorf$^{1,7 \ast}$ \and
	\small$^{1}$Faculty of Physics and CENIDE, University of Duisburg-Essen, 47048 Duisburg, Germany.\and
  	\small$^{2}$current adress: Institute of Physical and Theoretical Chemistry
    University of Würzburg, 97074 Würzburg, Germany.\and  
    \small$^{3}$Kavli Institute of Nanoscience Delft, Delft University of Technology, Delft 2628 CJ, The Netherlands.\and
    \small$^{4}$current adress: QphoX, Delft 2628 XG, The Netherlands.\and
    \small$^{5}$4th Physics Institute and Research Center SCoPE, University of Stuttgart, 70569 Stuttgart, Germany \and
    \small$^{6}$School of Physics, University of Melbourne, Parkville, Victoria 3010, Australia   \and
    \small$^{7}$ICAN, University of Duisburg-Essen, 47048 Duisburg, Germany. \and
	\small$^\ast$Corresponding author. Email: karin.everschor-sitte@uni-due.de (K.E.-S.); meyerzh@uni-due.de (F.M.z.H.) \and
	\small$^\dagger$These authors contributed equally to this work.
}

\begin{document} 

\maketitle

\begin{abstract} \bfseries \boldmath
Topology describes properties of physical systems that remain constant under continuous deformations. For infinite vector waves, global topological invariants in position space are typically associated with periodic patterns. We demonstrate that even for aperiodic Helmholtz-decomposable wave fields, possessing only the wave's intrinsic periodicity, a topological invariant can be found in momentum space. This invariant, the linking number, represents a Berry phase. By utilizing electromagnetic and hydrodynamic surface waves, we confirm the robustness of the linking number against deformations, and experimentally observe discrete transitions between distinct topological sectors. The linking number captures the topology of vector wave fields across both continuous and discrete momentum spaces. Our work introduces a unified topological framework for vector wave fields, enabling their classification via a global invariant.
\end{abstract}

\noindent
The tremendous success of topology across the various areas of physics is based on its ability to describe complicated, intractable physical systems in terms of a few simple quantities: the topological invariants. These quantities are of utmost importance as they dictate the behavior of physical systems across all length and time scales. Once fixed, these topological invariants will not change under any continuous deformation, making physical properties that are directly related to them extremely attractive for technological applications\cite{Gilbert2021}.

Many topological invariants have been introduced, and each of them is specific to the properties of the underlying physical system. For instance, Chern-numbers and $\mathbb{Z}_2$-invariants in the bandstructure dictate the behavior of robust electronic transport in condensed matter systems\cite{Bernevig2025, Wen2017, Senthil2015, Lu2014}. Skyrmions occur in a variety of systems, including magnetism\cite{Bogdanov1994, Roessler2006, Muhlbauer2009, Fert2017,EverschorSitte2018}, optics\cite{Tsesses2018, Du2019, Davis2020, Gao2020, Shen2024a }, and hydrodynamics\cite{Smirnova2024, Wang2025}. The related Skyrmion number counts how many times a 3D vector field on a 2D plane in position space can be mapped onto a sphere. Skyrmion textures are accompanied by a vortex in the two-dimensional projection of this field, characterized by a winding number. Such vortices can also be found in free electron waves\cite{Verbeeck2010,McMorran2011}, flux tubes in type II superconductors \cite{Abrikosov1957, Essmann1967, Blatter1994}, hydrodynamics\cite{Bliokh2022, Bliokh2024, Smirnova2024}, and optical fields \cite{Nye1974, Allen1992, Beijersbergen1994,Kim2010, Yao2011, Spektor2017}.  

Applying topological concepts requires that the vectorial texture either approaches a uniform background for large distances (magnetic Skyrmion in a ferromagnetic background, c.f. Fig.~1a) or forms a periodic lattice.  It is the structural periodicity of photonic crystals and solid state systems that has made the description of the topology of such systems within their Brillouin zone so successful. Without structural periodicity or a uniform background, a local analysis within a chosen boundary is a commonly employed technique \cite{Du2019, Dai2020, Dai2022, Ghosh2023, Dreher2024}. Such local analysis neglects the infinite extension of the vector wave field and one must be aware that the calculated local topology will depend on the selected region and may not even be an integer. If no constant background exists and the vector field extends to infinity, the physical space is not compact, and a {\em global} topological invariant cannot be assigned. In other words: mapping a physical parameter space onto the order-parameter manifold that describes a possible topology of the field requires the physical space to be {\sl compactifiable}. 

Here we show that by considering the topology in momentum (Fourier) space rather than in position space, an appropriate topological invariant for infinitely extending vector wave fields can be defined. The here introduced momentum space linking number (\RSLN{}) $\mathcal{N}$ provides an intuitive topological description of two-dimensional vectorial wave fields that evade conventional position-space topological classification.  We derive how the \RSLN{} $\mathcal{N}$ relates to the geometric (or Berry) phase and discuss that the \RSLN{} relates to vortices within the vector wave field. Using experimentally-determined vector wave fields, we demonstrate the \RSLN{}'s robustness against spatial deformations, its temporal stability, and that it can be even applied to crystalline and quasicrystalline wave-fields with discrete momentum spaces.

% approx. 460 words. maybe shorten by 50?

\subsection*{Topology of Vectorial Wave Phenomena in Momentum Space}
The fundamental problem of assigning topological invariants to extending vector fields in position space is illustrated in Fig.~1. For the textbook case of an isolated Skyrmion in an otherwise uniform background it is straight-forward to define a smooth mapping of the vectors in position (base) space onto a (target) sphere of vector directions with a well-defined unique Skyrmion number $\mathcal{S}=1$ (Fig. 1A). The optical equivalent of such a Skyrmion, a target Skyrmion\cite{Tian2023} (Fig.~1B), already challenges this simple concept: it is composed of an infinitely-extending, periodic sequence of Skyrmions with alternating Skyrmion numbers $\mathcal{S}=\pm1$. It is impossible to globally map the vectors of such a field configuration onto a simple sphere and to unambiguously assign a Skyrmion number to it. Even worse, not all vector wave fields are as good natured as the target Skyrmion. While the exemplary three-dimensional vector wave field in Fig.~1C locally displays Skyrmion-like textures, it is riddled with locations of vanishing field intensity and undefined vector directions\cite{Berry2000, Berry2001}. Any mapping of such a field onto a sphere must inevitably leave holes, which makes a Skyrmion number not only ambiguous but fundamentally indeterminable.

\begin{figure}
    \centering
    \includegraphics[width=\linewidth]{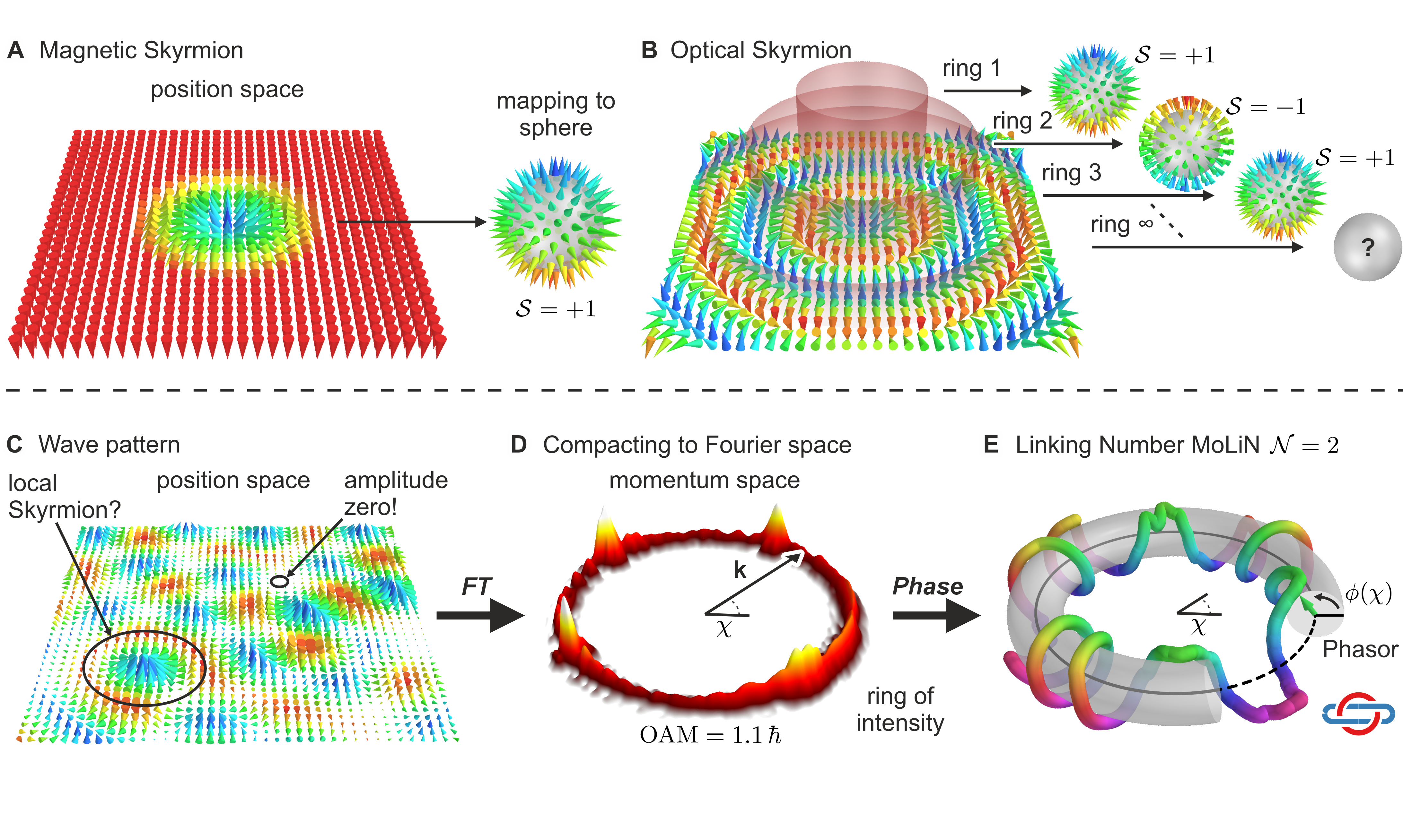}
    \caption{\textbf{Fundamental issue with characterizing the topology of extended wave fields in position space, and a solution using Fourier transformation.} (\textbf{A}) A magnetic Skyrmion is localized in space with field vectors that map onto a sphere. The mapping boundary is found from the regions where the orientations of the field vectors are fixed. (\textbf{B}) A target Skyrmion formed from an optical field has vectors that repeatedly oscillate in direction resulting in many possible mappings onto the sphere. The characterization of the topology depends on which boundary is chosen for the mapping. (\textbf{C}-\textbf{E}) show how a continuous optical field can be compacted in momentum space: (\textbf{C}) An example wave vector field with local Skyrmions and amplitude zeros. (\textbf{D}) In the Fourier transformed (FT) vector field the intensity is localized on a ring in momentum space. (\textbf{E}) Phasor $\phi(\chi) $ extracted from the vector field in momentum space winds around the momentum space ring. It is linked with the momentum space ring with a momentum space linking number (\RSLN ) of $\mathcal{N}=2$. Note that the expectation value for the orbital angular momentum (OAM) of this vector field does not coincide with the \RSLN .}
    \label{fig:figure_1}
\end{figure}

For an extended vectorial wave field in two dimensions a topological invariant can be found after Fourier-transforming the vectors $\mathbf{V}(\mathbf{r})$ from the position space with coordinate \textbf{r} to vectors $\tilde{\mathbf{V}}(\mathbf{k})$ in momentum space with momentum $\mathbf{k}$, as outlined in Figure~1D-E. For the monochromatic vector field of wavelength $\lambda$, like the one in Fig.~1C, the transformed vectors lie on a circle with radius $k_0=2\pi/\lambda$ in momentum space (c.f.~momentum intensity depicted in Fig.~1D). This circle captures the intrinsic periodicity of the vectorial wave field and defines a compact subregion of momentum space (\( S^1 \)-sphere), on which the vectors \( \tilde{\mathbf{V}}(\mathbf{k}) = \mathbf{\tilde{V}}(k_0 , \chi) \) depend only on the polar angle $\chi$. Note that for any vector field, including non-isotropic and non-monochromatic fields, the concept still applies as long as the fields transform into closed loops homotopic to $S^1$ in momentum space.

We will now show that the topology in Fourier-space is encoded by the winding of the vectors $\tilde{\mathbf{V}}(\chi)$ around the circle as function of  $\chi$. We achieve this by considering the geometric phase. Following Berry \cite{Berry1984} we express the Fourier-transformed field $\tilde{\mathbf{V}}(\chi)$ by a normalized polarization state $\tilde{\mathbf{V}}_\mathrm{pol}(\chi)$ and a phase $\phi(\chi)$: $\tilde{\mathbf{V}}(\chi)/\tilde{|\mathbf{V}}(\chi)| =  \ \tilde{\mathbf{V}}_\mathrm{pol}(\chi) \exp\left({i \ \phi(\chi)}\right)$. This phase describes the winding of the vectors $\tilde{\mathbf{V}}(\chi)$ around the circle and serves as a complete description of the topology of the complicated field. Overall, $\phi(\chi)\!: S^1\rightarrow S^1$ defines a mapping from a momentum space circle onto a circular target space. This mapping is visualized in Fig.~1E by drawing the phase $\phi(\chi)$ as a path on a torus, where $\chi$ is the toroidal direction and the phase $\phi(\chi)$ is mapped to the poloidal direction. The $S^1$ momentum space circle is represented by the black circle in the center of the torus, and the colored line represents the phase $\phi$ at each angle $\chi$. The phase curve winds around the torus, and if one tried to separate the colored curve in Fig.~1E from the black circle, this would be impossible because they are linked (twice), as depicted in the inset on the lower right of Fig.~1E. Such linking is described by the momentum space linking number \RSLN{} (here $\mathcal{N}=2$), which defines a topological invariant of this vector field within the homotopy group $\pi_1(S^1)$.

The \RSLN{} $\mathcal{N}$ can be understood in terms of the geometric phase acquired by the vector $\tilde{\mathbf{V}}(\chi)$ as one moves along the k-space circle. This phase is the Berry phase\cite{Berry1984, Bliokh2009} 
\begin{equation}
    \gamma = -i \int_0^{2\pi}  
    \tilde{\mathbf{V}}_\mathrm{pol}^\ast (\chi) e^{-i\phi(\chi)}
    \cdot \partial_\chi   
    \left(\tilde{\mathbf{V}}_\mathrm{pol}(\chi) e^{i\phi(\chi)}\right)
    \ \mathrm{d}\chi,
\end{equation}
\noindent\ and the \RSLN{} is $\mathcal{N}=\gamma/2\pi$. To fix the Berry-Phase to an integer multiple of $2\pi$, we require the polarization vector $ \tilde{\mathbf{V}}_\mathrm{pol}(\chi)$ to carry no geometric phase in itself (see Supplementary Text).

In the following we demonstrate experimentally the applicability of the \RSLN{} $\mathcal{N}$ and examine its stability against deformations, using the electromagnetic field vectors of surface plasmon polaritons and the displacement vectors of gravity-capillary water waves as examples. This selection of spin-momentum-locked waves exhibits spatial scales from nanometers to centimeters, time scales spanning from femtoseconds to milliseconds, and demonstrates the general applicability of our approach in two dimensions. The methods used to measure the different types of wave fields \cite{methods} are summarized in Fig. 2E-G. Time-resolved photoemission microscopy (TR-PEEM, Fig.~2E) and near-field scanning optical microscopy (NSOM, Fig.~2F) allow access to electric field vectors of surface plasmon polaritons.  The height-profile of water-waves and the related displacement vectors can be experimentally determined by the method of fast checkerboard decomposition\cite{Wildeman2018} (FCD, Fig.~2G).

\begin{figure}
    \centering
    \includegraphics[width=\linewidth]{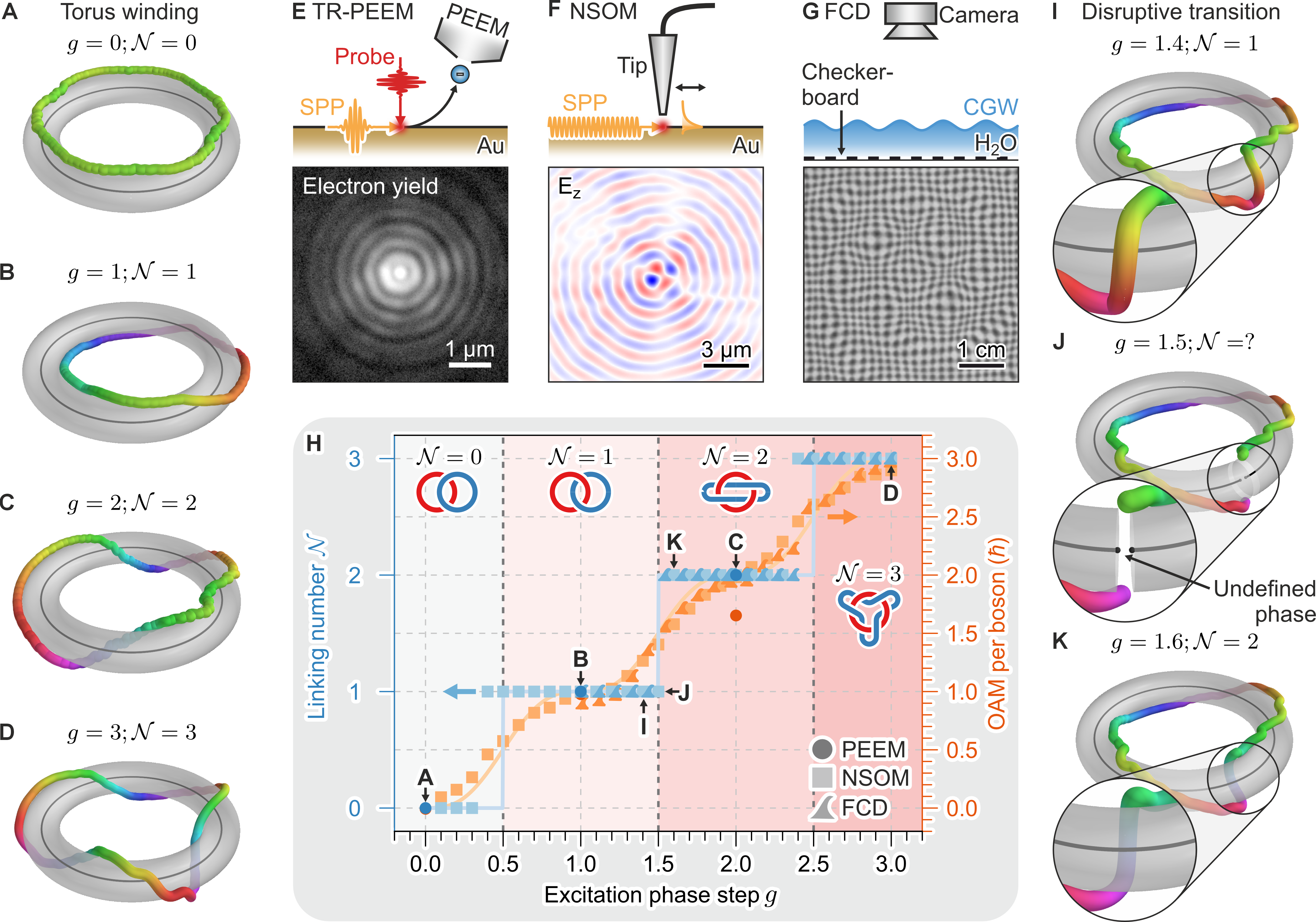}
    \caption{\textbf{OAM and momentum space linking number $\mathcal{N}$ of surface wave systems for various excitation phase steps $\textbf{g}$}. (\textbf{A}-\textbf{D})  Mapping of $S^1 \rightarrow S^1$ for integer excitation phase steps. (\textbf{H}) Direct comparison between \RSLN{} and OAM as a function of excitation phase step for the special case of superpositions of cylindrical harmonics analyzed here. The orange OAM data points follow a smooth line. The blue data points are linking numbers calculated from the same data. They follow a stepped curve as a function of excitation phase step. Within (H) data from different experimental methods is combined. These are: (\textbf{E}) Time-resolved photoemission microscopy; (\textbf{F}) Near field scanning optical microscopy; (\textbf{G}) Fast checkerboard decomposition of water waves. (\textbf{I}-\textbf{K}) Disruptive transition from the topological sector $\mathcal{N}=1$ to $\mathcal{N}=2$ of the mapping $S^1 \rightarrow S^1$ as a function of excitation phase step extracted from the NSOM data.}
    \label{fig:figure_2}
\end{figure}

To explore the concept of the \RSLN{}, the particular case of vector wave fields carrying orbital angular momentum (OAM) that can be described by a superposition of cylindrical harmonics \cite{Bauer2023} provide an excellent testbed. For the here considered surface waves, the OAM manifests as intensity-weighted phase vortices in the out-of-plane component of the field \cite{Berry2022,Davis2023}. Usually, the topological charge of such a vortex is defined as the winding number of the phase on a closed loop around the core of that vortex. Remarkably, the OAM expectation value of a field carrying integer OAM \cite{Allen1992,Simpson1997,Leach2004, Bauer2023} coincides with the locally-computed topological charge of the central vortex. In comparison, the \RSLN{} directly provides a global measure of the topology of the entire field by collapsing it onto a single phasor trajectory in momentum space. Figs.~2A-D show experimentally-determined phasor trajectories $\phi(\chi)$ with topological vortex charges of zero to three, respectively. As is evident from the winding of the phasor trajectories around the torus, the \RSLN{} $\mathcal{N}$ is identical to the topological charge of the central vortex of the fields, and is thus consistent with the OAM expectation value (see Supplementary Text for analytical proof).

The situation is more complicated for fields carrying fractional OAM, where the central single isolated vortex breaks up into an infinitely-extending sequence of vortices and anti-vortices\cite{Berry2004,Leach2004, Davis2023, Gbur2016, Bauer2023}. Such fields can be created by imprinting a phase discontinuity $2\pi g$ with an excitation phase-step $g$ into the boundary condition for the azimuthal phase dependence \cite{methods}. For non-integer values of $g$, neither the winding number in real-space nor the OAM expectation value over the whole field serve as an unambiguous topological classification of the field. First, any finite closed loop in position space will exclude an infinite subset of vortices located outside the loop, leaving the computed winding number loop-dependent\cite{Berry2004, Davis2023, Goette2008}. 
Second,  the OAM expectation value shows a continuous nonlinear dependence on the excitation phase step $g$ (orange curve and data points in Fig.~2H, right ordinate) \cite{Leach2004, Bauer2023}, disqualifying it as a topological invariant. Both issues, the infinite number of vortices and the continuously changing OAM values, are resolved by the here introduced \RSLN{} $\mathcal{N}$. The blue curve and data points in Fig.~2H (left ordinate) plot the calculated \RSLN{}  $\mathcal{N}$ for the same data used to calculate the OAM. Even though OAM and \RSLN{} appear related in the present case, there is no general relationship between OAM, phase vortices \cite{Berry2022}, and the \RSLN{} $\mathcal{N}$.

The \RSLN{} $\mathcal{N}$ is always an integer and changes whenever the excitation phase step crosses half integer values, i.e., $\mathcal{N}=[g]$. It is the integer nature of the \RSLN{}  $\mathcal{N}$ that rigorously defines distinct topological homotopy classes of vector fields, enabling a concise topological classification of extended wave phenomena. All field configurations with the same \RSLN{} are homotopic to a particular $ \mathcal{N}$-link as indicated by the unlink, single, double, and triple link icons at the top of Fig.~2H. The NSOM data in Figs.~2I-K displays in detail how the linking changes when $g$ is gradually increased across a half-integer value. For $g=1.4$ (Fig.~2I) the phasor curve wraps around the torus once, whereas for $g=1.6$ (Fig.~2K) it wraps around the torus twice. For $g=1.5$ (Fig.~2J) the phasor curve is undefined in exactly one point, which can be interpreted as a topological phase transition that mediates a discontinuous transition between two adjacent topological sectors. In position space, this transition can be described by adding a single vortex coming {\em from infinity} to the already infinite series of vortices and anti-vortices  in position space\cite{Gbur2016, Davis2023, Bauer2023}.

\subsection*{Stability of the Linking Against Deformation}
\begin{figure}
    \centering
    \includegraphics[width=\linewidth]{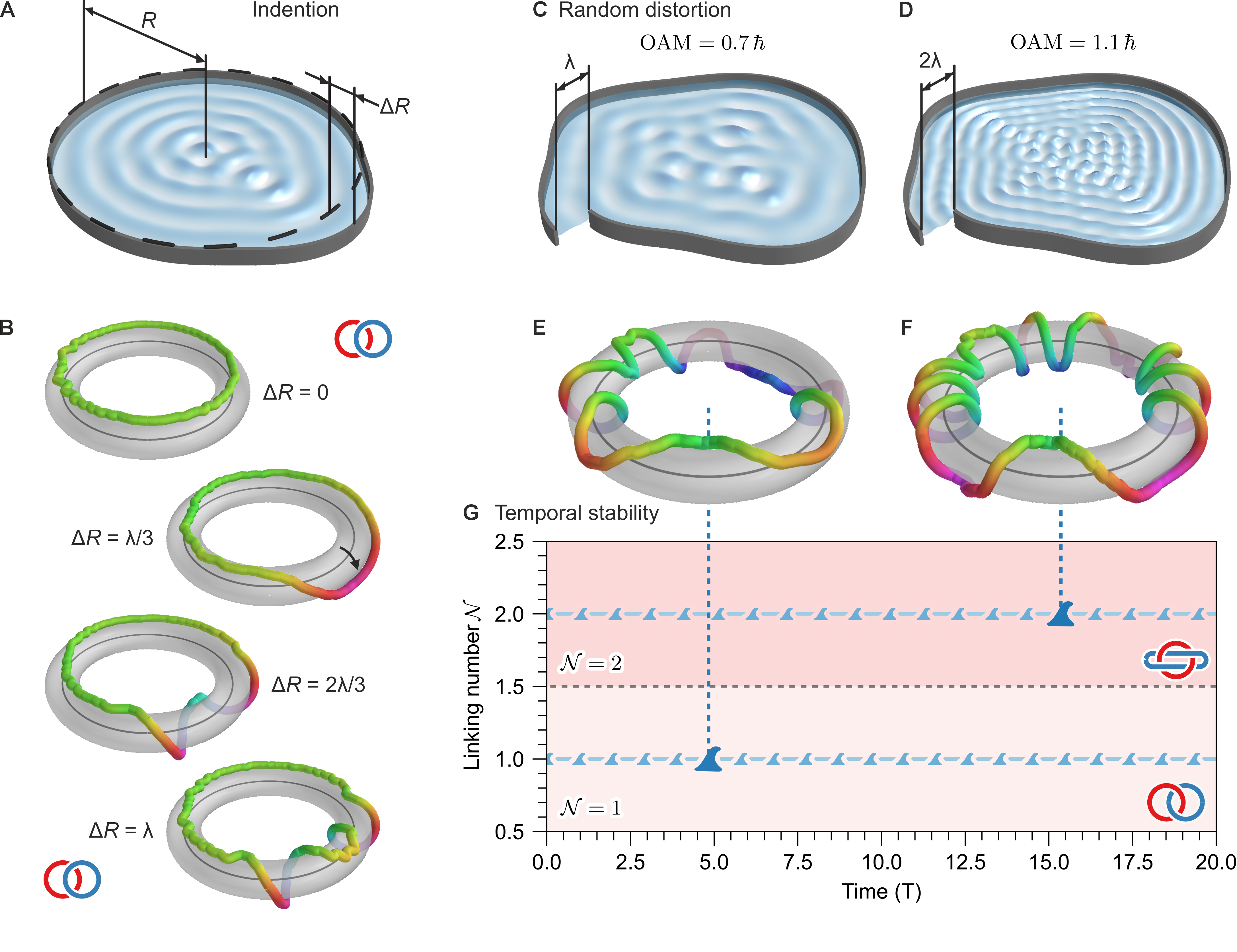}
    \caption{\textbf{Stability against deformations.} (\textbf{A}) A water wave obtained from experiments using a deformed circular excitation structure. (\textbf{B}) Corresponding phasor winding plots as a function of the deformation $\Delta R$. (\textbf{C} and \textbf{D}) A strongly deformed Archimedean spiral excitation structure is used to excite the measured water wave patterns with excitation phase step $g=1$ and $g=2$, respectively. (\textbf{E} and \textbf{F}) Corresponding phasor windings calculated from the experiments are drawn in (C) and (D). (\textbf{G}) The temporal evolution of the \RSLN{} $\mathcal{N}$ , extracted from experimental data presented in (E) and (F). T on the time axis denotes full oscillation periods. For clarity, only every $20^{\mathrm{th}}$ ($10^{\mathrm{th}}$) datapoint is shown for $\mathcal{N}=1$ ($\mathcal{N}=2$)}
    \label{fig:figure_3}
\end{figure}

For the \RSLN{} to be a  topological invariant, it must remain robust against any continuous deformation of the system. A simple deformation of the vector field can, for example, be accomplished by a smooth indentation of an otherwise circular excitation structure, as shown using water waves in Fig.~3A. The initially undeformed wave excitation structure is a circle of radius $R$  (dashed line). At the lower right of the circle a continuous protrusion with dimension $\Delta R$ is introduced, which changes the water wave landscape (blue-shaded, experimentally-reconstructed water wave surface in Fig. 3A). The corresponding evolution of the phasor trajectory on the torus as function of $\Delta R$ is displayed in Fig.~3B. At $\Delta R=0$, the structure resembles a simple circle. Therefore, the wave pattern exhibits no vortices and reproduces the expected unlink (c.f. Fig.~2A). Increasing the protrusion $\Delta R$ to $\lambda/3$ not only creates vortex/anti-vortex pairs in position space, but it also initiates an outward winding of the phasor around the torus in momentum space. It is important to note that, while the phasor curve is not a simple circle any more, its partial winding around the torus on the lower right of Fig.~3B ($\Delta R = \lambda/3$) is compensated by a corresponding unwinding to maintain the unlink.
If $\Delta R$ is increased further, the local winding of the curve becomes more pronounced, but remains compensated by the corresponding unwinding, so that the overall \RSLN{}  $\mathcal{N}$ remains invariant.

While this simple example illustrates the response of the phasor curve to a local protrusion, the concept can be carried over to extended deformations of the entire excitation phase landscape. Figure~3C depicts a water wave surface excited by a heavily deformed Archimedean spiral \cite{methods} with an excitation phase step of $g=1$. The deformation makes the height profile of the water wave highly irregular and rich with vortices and anti-vortices. The corresponding phasor trajectory in Fig.~3E exhibits a complicated winding that can mostly be decomposed into pairs of winding and unwinding like it was observed for the simple protrusion in Figs.~3A,B. The remaining winding is solely due to the excitation phase step $g = 1$. Despite the deformation, the wave retains a \RSLN{} of $\mathcal{N} = 1$, and thus belongs to the same homotopy class as the prototypical OAM field in Fig.~2B. Remarkably, the OAM expectation value of this vector field differs from the \RSLN{}, even though the field is excited with an integer phase step. If the wavelength of the water wave is halved, the same deformed excitation structure exhibits a phase step of $g=2$ (Fig.~3D). The corresponding phasor curve in Fig.~3F again rapidly winds and unwinds around the torus. It nevertheless results in an overall \RSLN{} of $\mathcal{N}=2$ and is homotopic to that of the OAM field in Fig.~2C. It would be challenging to arrive at a topological classification of such a field from its vectorial structure in position space, whereas the computation of the \RSLN{} $\mathcal{N}$ is global and straight-forward. While the experiments were limited to specific deformations of the boundary, it is analytically derived in the Supplementary Text that the \RSLN{} $\mathcal{N}$ is invariant against any continuous deformation of the boundary irrespective of its specific nature.

Any topological invariant must also be temporally stable as long as no disruptive perturbation is introduced in time. From time-resolved recordings of the water wave configurations in Figs.~3E,~F, we plot the time dependence of the respective \RSLN{}s in Fig.~3G. As expected the \RSLN{}s do not change during the time of observation (20 oscillation periods T).
 
\subsection*{Topology in Discrete Spaces}
Until now, the discussed wave phenomena were composed of interference fields where the momentum vectors formed a closed continuous loop in momentum space. Wave systems ranging from optical lattices to Bloch waves and magnetic Skyrmion crystals, however, commonly exhibit spatial order that yield discrete momentum space representations. 
These are no longer a part of the homotopy group $\pi_1(S^1)$, as they do not form a closed loop in momentum space. In the following, we will demonstrate that the \RSLN{} intrinsically captures discrete momentum spaces as well, using experiments with discrete numbers of interfering waves. For instance, the momentum space of the interference of six plane water waves exhibits $M=6$ well-defined intensity peaks (bottom of Fig.~4A). If arranged according to their discrete angle $\chi_i$ in momentum space, the momentum  points form a cyclic graph $C_M$. For each of these points the phase $\phi_i=\phi(\chi_i)$ is extracted and displayed as a phasor vector in the center of Fig.~4A. These phasor vectors perform a cumulative rotation of two full turns, establishing a \RSLN{} of $\mathcal{N}=2$. This becomes evident via construction of a phasor trajectory on the torus by geodesic interpolation in between the discrete phasor vectors (top of Fig. 4A).

\begin{figure}
    \centering
    \includegraphics[width=\linewidth]{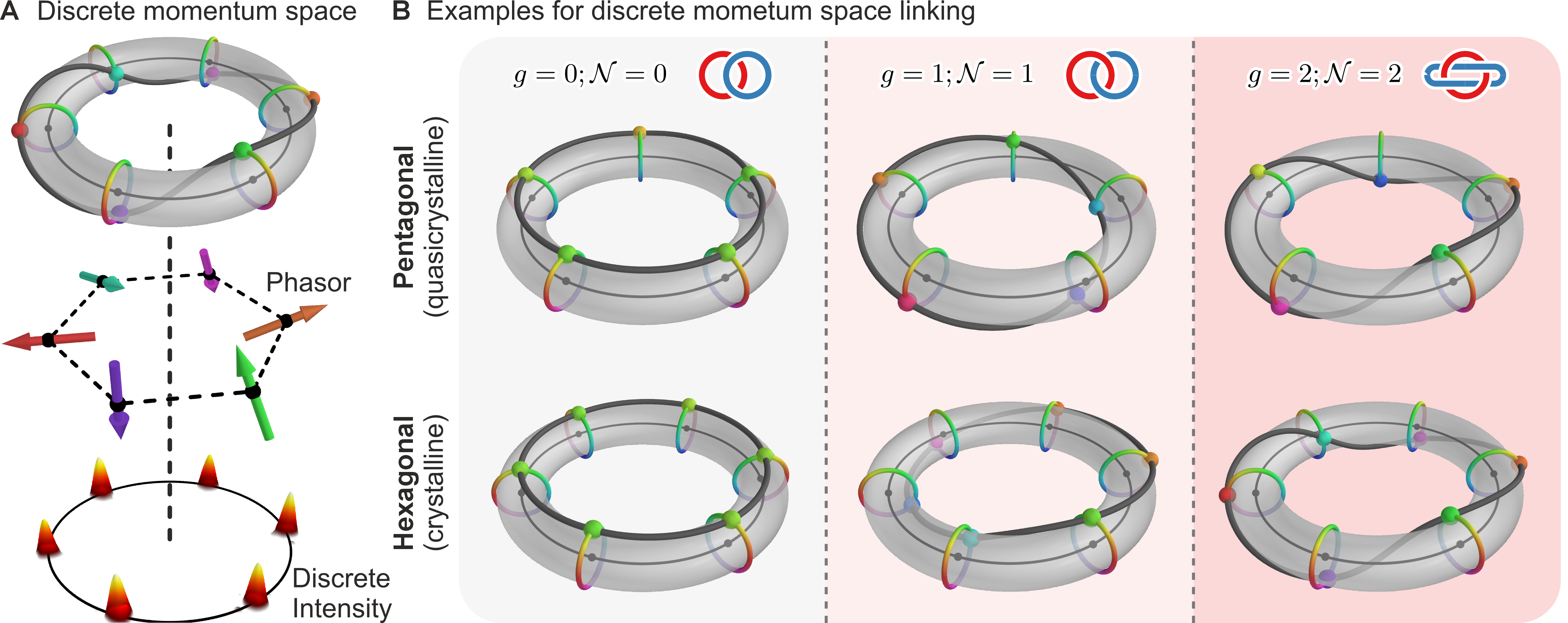}
    \caption{\textbf{Topology on discrete momentum spaces.} (\textbf{A}) Discrete momentum space leads to a generalization of the linking number. By connecting the phasors at the discrete peaks on the torus, a linking curve is constructed. (\textbf{B}) Examples of discrete momentum spaces with different linking numbers of quasicrystalline and crystalline vector wave fields.}
    \label{fig:figure_4}
\end{figure}

Figure~4B extends this framework to discrete momentum spaces with pentagonal (quasicrystalline) symmetry ($M=5$, top row) and hexagonal (crystalline) symmetry ($M=6$, bottom row). The hexagonal momentum space with $g=0$ (left column, bottom row of Fig. 4b) is calculated from the data used by Davis et al.\cite{Davis2020} to demonstrate the temporal dynamics of a plasmonic Skyrmion lattice. The cases for $g=1$ and $g=2$ are calculated from water wave data and produce \RSLN{}s of $\mathcal{N}=1$ and $\mathcal{N}=2$, respectively.

The pentagonal momentum space with $g=0$ (left column, top row) is calculated from the plasmonic quasicrystal data used by Tsesses et al.\cite{Tsesses2025}.  If we add an excitation phase step $g\neq 0$, the extracted \RSLN{}s  increase to $\mathcal{N}=1$  for $g=1$ (center column) and to $\mathcal{N}=2$ for $g=2$ (right column), here shown for water waves. As the topological charge vectors are only defined by the relative phases of the five waves forming the quasicrystal, the \RSLN{} directly measures the same topological invariant that was derived from topological charge vectors in four dimensions\cite{Tsesses2025}. 
Even though position space and momentum space are both strictly aperiodic, the \RSLN{} still captures the correct topology.

A special configuration in discrete momentum spaces arises for $g=M/2$. For instance, for $M=6$ and $g=3$ adjacent phasors are anti-parallel, rendering the handedness of the winding between them ambiguous. The resulting wave-field, despite $g=3$, does not carry any orbital angular momentum, and as such maps to an \RSLN{} of $\mathcal{N}=0$. Generally, such aliasing appears for excitation phase steps $|g|\geq M/2$, since $M/2$ is the Nyquist frequency of the cyclic graph $C_M$. The homotopy group for discrete momentum spaces can thus be identified with a subset of $\mathbb{Z}$, restricting the possible \RSLN{}s to a range $-M/2<\mathcal{N} < M/2$. 

\subsection*{Conclusions}
For infinitely extending vector fields the position space is typically not compact and as a consequence, it is not possible to properly define a global topological invariant in position space.  
Here, we use a Fourier transform to create a compact manifold in momentum space. Infinitely extending monochromatic two-dimensional vector wave fields can be classified by the momentum space linking number (\RSLN{}) $\mathcal{N}$.
While the \RSLN{} is a global topological invariant, it can be
experimentally inferred from a measurement of local field vectors over a finite measurement region in position
space. Importantly, the finite extent of this region only restricts the resolution of the accessible momentum space.
The \RSLN{} measures the winding of a phasor around a circle in momentum space and can be understood as a geometric (Berry) phase. It can be applied to periodic wave fields, but also describes aperiodic and quasicrystalline fields where no unit cell can be defined in position space. A generalized connection between the sum over all position space vortices and momentum space topology, based on the \RSLN{}, would be reminiscent of a Poincaré Hopf theorem for non-compact manifolds \cite{Dreher2024, Bucher2026}. The \RSLN{} $\mathcal{N}$ is then an appropriate topological invariant in situations with many or an infinite number of vortices. 

%This allows one to provide a measure of topology on much more complex topology. it extends the toolbox of topology to globally describe truly unintuitively complicated vectorial fields, Quasicrystals and arrives at a simple integer number that is robust against deformation

Our framework establishes a global topological measure for highly complex vector fields. It extends topological methods to capture otherwise unintuitive structures and reduces their complexity to a single integer invariant that remains stable under continuous deformations. As such, the description of many wave-field phenomena will benefit from the \RSLN{}. For instance, in periodic spin-momentum-locked wave fields, vanishing \RSLN{} allows for non-trivial skyrmionic topology per unit cell \cite{Tsesses2018,Davis2020,Smirnova2024,Schwab2025,Wang2025}.

%, whereas for \RSLN{} $\mathcal{N}=1$ the corresponding texture is instead found in the spin angular momentum\cite{Ghosh2021, Schwab2025}.

The intrinsic robustness of the \RSLN{} makes it a promising candidate for data storage and transfer applications \cite{Shen2024a, Guo2026}, much like the robustness of topologically protected modes has been used to obtain scattering-free transport\cite{Wang2009, Mei2019}. Especially in data transmission using OAM modes \cite{Mair2001, HamadouIbrahim2013, Trichili2016, Guo2026}, one should reconsider whether the \RSLN{} $\mathcal{N}$ is a better carrier of topologically protected information.

The \RSLN{} provides a common topological language for monochromatic Helmholtz wave systems across optical, acoustic, elastic, plasmonic, and matter-wave platforms. By decoupling topology from microscopic structure and periodicity, it establishes a general framework for extended wave fields and opens a route from topological classification to function.

\clearpage 

\bibliography{Bibliography} % for a file named science_template.bib
\bibliographystyle{sciencemag}

%%%%%%%%%%%%%%%% ACKNOWLEDGEMENTS %%%%%%%%%%%%%%%

\subsection*{Acknowledgments}
We thank Julian Schwaab and Laurens Kuipers for discussions and valuable input.

\paragraph*{Funding:} We acknowledge support from the ERC (grant no. 321331-Complexplas (B.F. and H.G.), 
grant no. 862549-3DPrintedoptics (B.F. and H.G.), grant no. 340438-CONSTANS (Th.B.)), DFG 
(grant no. SPP1391 Ultrafast Nanooptics (A.N., P.D., D.J., A.R., M.M., B.F., H.G., and F.-J.M.z.H.), CRC 1242 Non-Equilibrium Dynamics of Condensed Matter in the Time Domain project no. 278162697-SFB 1242 (A.N., P.G., P.D., D.J., A.R., M.M., M.A., K.E.-S. and F.-J.M.z.H.)), BMBF (Printoptics (B.F. and H.G.), BW Stiftung (Spitzenforschung (B.F. and H.G.), Opterial (B.F. and H.G.)) and Carl-Zeiss Stiftung (B.F. and H.G.). T.J.D. acknowledges support from the MPI Guest Professorship Program and from the DFG (grant no. GRK2642 Photonic Quantum Engineers) for a Mercator Fellowship. M.A. also acknowledges support from the UDE Postdoc Seed Funding Project TORUS.

\paragraph*{Author contributions:}
Conceptualization: A.N., P.G., P.D., K.E.-S., and F.M.z.H.  

Methodology: A.N., P.G., P.D., M.A., and T.J.D.  

Software: A.N. and P.D.  

Resources: B.F. and H.G. fabricated the samples for the polarimetric PEEM measurements.  

Investigation: A.N., P.D., D.J., A.R., M.M., and F.M.z.H. performed the polarimetric PEEM experiments; Th.B. performed the NSOM experiments; A.N. designed and performed the water-wave experiment.  

Validation: A.N., P.D., Th.B., and F.M.z.H.  

Formal analysis: A.N.  

Visualization: A.N., M.M. and T.J.D.  

Writing – original draft: A.N., P.G., P.D., T.J.D., F.M.z.H.  

Writing – review \& editing: all authors.  
Supervision: H.G., K.E.-S., and F.M.z.H.  
Funding acquisition: H.G., K.E.-S., and F.M.z.H.

\paragraph*{Competing interests:}
The authors declare no competing interests.

\paragraph*{Data and materials availability:}

All data are available in the manuscript or the supplementary materials.

\subsection*{Supplementary materials}
Materials and Methods\\
Supplementary Text\\
Figs. S1 to S3\\
References \textit{(60-\arabic{enumiv})}\\ % automatically fills out the last reference number
% (filling out the other numbers automatically is possible but fiddly and liable to break)
Movie S1-S4\\

\newpage
\renewcommand{\thefigure}{S\arabic{figure}}
\renewcommand{\thetable}{S\arabic{table}}
\renewcommand{\theequation}{S\arabic{equation}}
\renewcommand{\thepage}{S\arabic{page}}
\setcounter{figure}{0}
\setcounter{table}{0}
\setcounter{equation}{0}
\setcounter{page}{1} % not 0 as \newpage already started a supplementary page
% References continue the numbering from the main text.

\begin{center}
\section*{Supplementary Materials for \\ \scititle}
A.~Neuhaus$^{\dagger}$,
P.~Gessler$^{\dagger}$,
P.~Dreher$^{\dagger}$,
D.~Janoschka, 
A.~R\"odl, 
M.~Manten, 
Th.~Bauer, 
M.~Azhar,
B.~Frank, 
T.J.~Davis, 
H.~Giessen, 
K.~Everschor-Sitte$^{\ast}$, 
F.~Meyer~zu~Heringdorf$^{\ast}$ \\
% Identify at least one corresponding author, with contact email address
\small$^\ast$Corresponding author. Email: karin.everschor-sitte@uni-due.de ~(K.E.-S.); meyerzh@uni-due.de~(F.M.z.H.) \\
% Joint contributions can be indicated like this
\small$^\dagger$These authors contributed equally to this work.

\end{center}
\subsubsection*{This PDF file includes:}
Materials and Methods\\
Supplementary Text\\
Figures S1 to S3\\
Captions for Movies S1 to S4\\

\subsubsection*{Other Supplementary Materials for this manuscript:}
Movies S1 to S4\\

\newpage

%%%%%%%%%%%%%%%% MATERIALS AND METHODS %%%%%%%%%%%%%%%

\subsection*{Materials and Methods}

\subsubsection*{Polarimetric PEEM}
The spatio-temporal evolution of surface plasmon polariton (SPP) fields was reconstructed using time-resolved polarimetric two-photon photoemission (2PPE) electron microscopy (PEEM)\cite{Davis2020, Dreher2024}. A femtosecond laser pulse, normally incident on a single-crystal Au(111) platelet\cite{Radha2010,Radha2011}, excited SPPs at grooves patterned by focused ion-beam milling (Raith ionLINE Plus) that provided the necessary in-plane momentum. The groove geometry and the polarization of the excitation determined the resulting field configuration, with both continuous (Archimedean spirals, $L = 1; 2 $, helicity $S = \pm1$)\cite{Kim2010, Spektor2017, Spektor2019} and discrete (hexagonal\cite{Tsesses2018, Davis2020} and pentagonal\cite{Tsesses2025}) coupling structures employed. The excitation pulse had a central wavelength of 800 nm and duration below 15 fs, generating fs pulses of long-range SPPs with a central wavelength of $\lambda_{\mathrm{SPP}} \approx 780 ~\mathrm{nm}$. A delayed phase-stable pulse with tunable polarization, derived from a Mach–Zehnder interferometer with attosecond stability\cite{Neuhaus2026}, probed the evolving SPP field after a variable delay $\Delta \tau$. The interference between the SPP-pulse and the probe pulse initiated two-photon photoemission\cite{Kubo2007,Chelaru2007}, and the emitted electrons were subsequently imaged in an ELMITEC SPELEEM 3 microscope\cite{Schmidt1998}. Measurements were repeated over multiple polarizations and delay times. As the electron detector attached to the microscope \cite{Janoschka2021} is much slower than the time scales of a single pump-probe experiment, it averages over multiple experiments under the same conditions. We employ a mode-locked fs-laser with a 80 MHz repetition rate to achieve statistical convergence. The resulting datasets recorded the spatially resolved electron yield as a function of probe polarization and delay, yielding a for dimensional dataset.

Subsequent data processing involved drift correction, delay calibration and Fourier filtering to isolate the fundamental SPP frequency and momentum components. By combining the different probe polarizations the in-plane electric field vectors were reconstructed directly, while the out-of-plane component was computed using Maxwell’s equations and the evanescent decay of the near-field into the vacuum. A home-built numerical fixed-point iteration scheme\cite{Dreher2024} corrected the nonlinear PEEM response to ensure correct field reconstruction under strong focussing conditions, \cite{Dreher2022}. More details on the data processing can be found in \cite{Dreher2024}. This methodology yields a complete vectorial and time-resolved reconstruction of the electric near-field at the surface, enabling the direct observation and analysis of topological field configurations in plasmonic wave systems.

\subsubsection*{Nearfield Scanning Optical Microscopy}
The Near-field Scanning Optical Microscopy experiments were performed in a home-built phase-and polarization-resolving near-field microscope. A continuous wave excitation light from a SANTEC TSL-710 laser at $\lambda=1550\,\mathrm{nm}$ was split into a signal and reference branch for interferometric phase detection. The signal light was spatially tailored in its polarization and phase via a spatial light modulator (Meadowlark Optics P1920-1100-1550) in double-pass configuration to transform the Gaussian  beam shape into a radially polarized mode with annular intensity distribution and an azimuthal phase ramp of $2\pi g$. The tailored light field illuminated a circular slit of radius $75 \,\mathrm{\upmu m}$ in a $200\,\mathrm{nm}$ thick gold film, fabricated via focused ion beam milling. 

The excited surface plasmon polariton field was collected through an aperture-based near-field probe consisting of a tapered single-mode optical fiber coated with $180\,\mathrm{nm}$ aluminum and an aperture of $200\,\mathrm{nm}$ at its apex, raster-scanned over the surface of the gold film at a height of ca. $20\,\mathrm{nm}$, controlled through shear-force feedback. We separately detected the two orthogonal in-plane polarization components of the SPP field via a polarization-resolved heterodyne scheme to extract the full vectorial in-plane field information. The z-component of the SPP field was then unambiguously reconstructed from this field information due to the pure TM nature of the SPPs. More details on the experimental setup and measurement method can be found in references \cite{Burresi2009, Bauer2023}.

\subsubsection*{Water Wave Experiment}

\begin{figure}[htbp]
    \centering
    \includegraphics[width=0.5\linewidth]{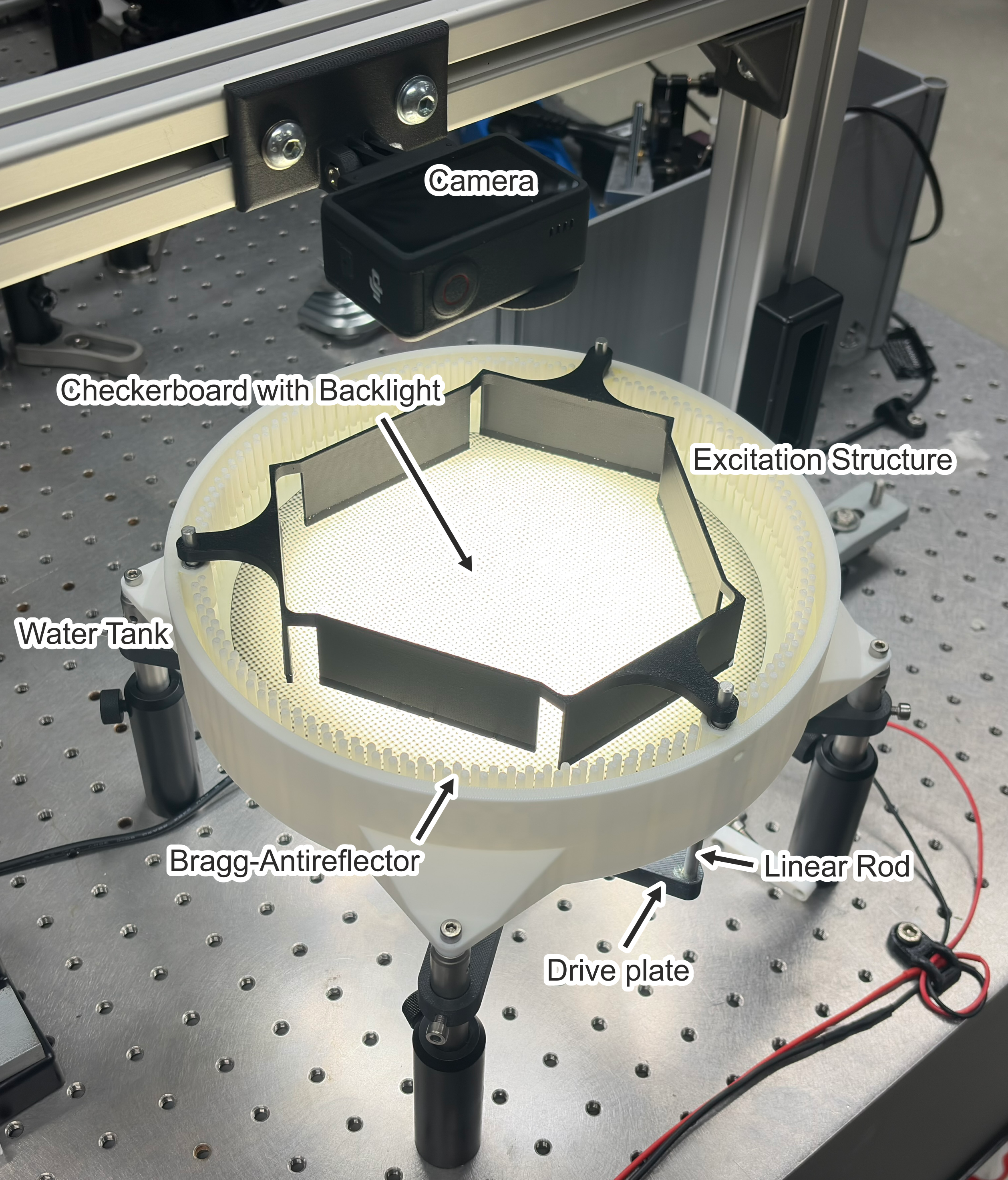}
    \caption{\textbf{Photograph of the experimental setup for the water wave experiments.} Multiple components are indicated. Water waves are launched from an oscillating excitation structure within the 3D printed water tank. A camera is placed on top of the tank to record videos of the distortions of the checkerboard in the water tank.   }
    \label{fig:Figure_S1}
\end{figure}

A photograph of the water wave experimental setup is depicted in Fig. \ref{fig:Figure_S1}. A circular 3D-printed water tank made of white PLA with a diameter of 230 mm and a water depth of 18 mm was used to study capillary–gravity surface waves near the transition regime. The tank’s outer wall integrated a Bragg-type wave absorber to suppress boundary reflections and maintain a clean central wave-field. The bottom surface featured a 2 mm-period sine-type two-dimensional checkerboard pattern, serving as an optical reference for fast checkerboard decomposition (FCD) \cite{Wildeman2018} and ensuring accurate height reconstruction and spatial calibration. Uniform LED back lighting beneath the semi-transparent tank provided homogeneous illumination and minimized Schlieren contrast.

Surface waves were generated by a 3D-printed excitation structure driven by a DC motor. The motor was connected to a drive plate below the water tank via an eccentric rod. The structure was connected to the drive plate via three rods guided by linear bushings to ensure purely vertical translational motion. Its geometry and drive frequency determined the excited wave pattern. For orbital angular momentum experiments, two Archimedean spirals with 20 mm and 40 mm gaps and inner diameters of 150 mm and 120 mm were used, respectively. Deformed circular and random spiral variants were also printed.  For the discrete configurations in the context of Fig.~4 of the main manuscript, the polygonal base structures were 3D-printed such that the individual wave sources were radially shifted outward according to their respective phase offsets. The system was driven at 12–25 Hz, corresponding to wavelengths between 10 mm and 20 mm. The vertical wave oscillation amplitude (200 µm) was small compared to the wavelength, ensuring linear wave behavior and precise reconstruction.

The surface was imaged using a DJI Action 5 camera (1920 × 1080 px, 240 fps, 1/500 s shutter, ISO 100, Normal Mode) mounted 145 mm above the tank bottom, aligned centrally to eliminate parallax. Optical distortion due to refraction was corrected using a geometric scale factor $f = 1.096$, yielding true lateral dimensions on the water surface. The FCD method reconstructed the surface height field $h(x,y,t)$, which was spatio-temporally Fourier-filtered to extract the fundamental mode. From this, the in-plane displacement field $\mathbf{u}(x,y,t)=\frac{i}{k_0}\mathbf{\nabla} h(x,y,t)$ was derived for quantitative vector wave field analysis.

\subsubsection*{Imprinting Excitation Phase-steps}
Many of the experiments discussed in the main manuscript rely on specific excitation geometries to imprint a phase excitation step $g$ on the generated waves. For each of the used experiments, the excitation phase step is imprinted in a different manner. For the time-resolved PEEM experiment, the excitation is created at a boundary in the shape of  an Archimedean spiral, using circularly polarized light for illumination \cite{Kim2010, Spektor2017}. In the NSOM experiment, the phase step $g$ is realized by exciting the wave at a circular boundary with radially polarized light that had a phase-step imprinted by a spatial light modulator\cite{Bauer2023}. In the water wave experiment the excitation is caused by a boundary at the top of the water surface that is continuously moved up and down, as described above.  The excitation phase step can then either be controlled by changing the shape of the boundary, or by changing the driving frequency of the excitation.

\subsubsection*{Deformed Excitation Coupler} \label{sec:DeformedExcitationCoupler}
This note provides additional experimental details on the three-dimensional printed spiral structure used in Fig.~3C-G of the main manuscript. The construction of the structure is illustrated in Fig. S2. The structure is based on an Archimedean spiral with an inner diameter of $75\,\mathrm{mm}$ and a gap of $20\,\mathrm{mm}$ that is drawn as a dashed line in Fig.~S2. The wall thickness is $2\,\mathrm{mm}$. Along the reference Archimedean spiral, six base points are selected (blue X's). Two of these points are located at the ends of the gap and are assigned zero radial displacement. At the remaining four points a random radial displacement is applied. The displacement is drawn from a uniform distribution of width $15\,\mathrm{mm}$ and is defined with respect to the local radius of the Archimedean spiral. The radial displacement values at all six points, including the zeros at the gap, are interpolated as a function of the polar angle using a cubic periodic spline. This yields a continuous radial displacement as a function of polar angle that lies within a deformation corridor of total width $27\,\mathrm{mm}$ around the reference spiral (gray shaded area in Fig.~S2). Adding this interpolated radial displacement to the radius of the Archimedean spiral defines the final deformed structure, drawn as a solid black line in Fig.~S2. Using this structure to excite water waves with wavelengths of $10~\mathrm{mm}$ and $20~\mathrm{mm}$ corresponds to excitation phase steps of $g = 2$ and $g = 1$, respectively.

\subsection*{Supplementary Text}
\subsubsection*{Geometric Phase and Linking}
This section of the supplementary text will include the analytical derivations showing under what conditions the \RSLN{} is a topological invariant, and how it connects to the geometric (or Berry) phase of the field.
To this extent, the first part of this section discusses how the momentum space linking number \RSLN{} can be understood as the acquired geometric phase of the vector field  $\tilde{\mathbf{V}}(\mathbf{k)}$ along the momentum space ring, as discussed in the main manuscript. We derive which conditions the vector field needs to fulfill to result in the \RSLN{} as the topological invariant of the system.
The equivalence between the geometric phase and the linking between the momentum space circle and the phasor curve is demonstrated within the second part of this section \ref{sec:Linking_Phase} using a geometric argument.
The third part derives analytically that there is no general relation between orbital angular momentum and linking number, and that in the special case of pure cylindrical harmonics these coincide for integer excitation phase steps.
Lastly, we analytically show the invariance of the \RSLN{} against any continuous deformation of the excitation boundary.

\paragraph*{Geometric Phase}
\label{sec:Geometric_Phase}
Here we will derive the conditions of a vector wave field under which the \RSLN{} $\mathcal{N}$ becomes a topological invariant.
Many physical systems, such as acoustical, electromagnetic, elastodynamic, quantum, and even harmonically driven diffusive (heat) systems, reduce to the Helmholtz partial differential equation under a time-harmonic ansatz, we consider a vector field $\mathbf{V}(\mathbf r)$ that obeys the Helmholtz equation $\nabla^2 \mathbf V(\mathbf r)+k_0^2 \mathbf V(\mathbf r)=0$. %We also impose the condition that the vector is source free so that $\nabla \cdot \mathbf V(\mathbf r)=0$. 
We write the spatial Fourier transform of the vector field as $\tilde {\mathbf V}(\mathbf k)$. In general this vector $\tilde {\mathbf V}(\mathbf k)$ is complex and lies on the surface of a sphere with fixed radius $k_0$. For vector directions $\mathbf k$ confined to two dimensions, this becomes a ring of radius $k_0$ in the plane, so that the vector components are only functions of the angle $\chi$, representing the orientation in Fourier space. %In other words, the vector field is orthogonal to the propagation direction, and we assume it remains so under slow variations of the parameters. This is equivalent to the adiabatic condition used to describe geometric phases \cite{Berry1984}.
Without loss of generality we can extract a global phase $\phi(\chi)$ from the vector components so that the normalized vector field $\tilde{\mathbf{V}}(\chi)/\tilde{|\mathbf{V}}(\chi)| =  \ \tilde{\mathbf{V}}_\mathrm{pol}(\chi) \exp\left({i \, \phi(\chi)}\right)$, where $\tilde {\mathbf V}_\mathrm{pol}(\chi)$ is the normalized polarization vector as used in the main manuscript, i.e. $\tilde{\mathbf  V}_\mathrm{pol}^* \cdot \tilde{\mathbf V}_\mathrm{pol}=1$. Note that $\tilde{\mathbf V}_\mathrm{pol}$ may still be complex. Consider the following integral, which is similar to the adiabatic transport equation derived by Berry \cite{Berry1984}
\begin{equation}
\label{eq:bp}
\begin{split}
\gamma&=-i\int_0^{2\pi} \tilde{\mathbf V}_\mathrm{pol}^* \, e^{-i \phi} \cdot \frac{\partial}{\partial \chi} \left (\tilde{\mathbf V}_\mathrm{pol} \, e^{i \phi} \right )  \mathrm{d}\chi\\
&=-i\int_0^{2\pi} \left (i \frac{\partial \phi}{\partial \chi} +\tilde {\mathbf V}_\mathrm{pol}^* \cdot \frac{\partial \tilde {\mathbf V}_\mathrm{pol}}{\partial \chi} \right ) \mathrm{d}\chi.
\end{split}
\end{equation}
This integral has the form of the adiabatic variation of the field under changes to the parameter $\chi$, being the angle in Fourier space.
Under the condition that the polarization vector $\tilde{\mathbf V}_\mathrm{pol}$ itself carries no geometric phase, i. e., 
\begin{equation}
\label{eq:zero}
\int_0^{2\pi} \tilde {\mathbf V}_\mathrm{pol}^* \cdot \frac{\partial \tilde {\mathbf V}_\mathrm{pol}}{\partial \chi}  \mathrm{d}\chi=0,
\end{equation}
then Eq. \eqref{eq:bp} becomes the geometric phase of the Fourier transform of the field %, also known as the Berry phase \cite{Berry1984}
\begin{align}
\label{eq:gemoetric_phase}
\gamma&=\int_0^{2\pi} \frac{\partial \phi}{\partial \chi} \mathrm{d}\chi \\
\mathcal{N}&=\gamma/2\pi,
\end{align}
with the momentum space linking number (\RSLN{}) $\mathcal{N}$ as introduced in the maintext. Spin-momentum locked waves have ${\tilde {\mathbf V}_\mathrm{pol}^*\cdot \partial _\chi \tilde{\mathbf V}_\mathrm{pol} =0}$ at all points on the ring in Fourier space so that condition Eq. \eqref{eq:zero} always holds. In all our experiments we use spin-momentum locked waves. 

\paragraph*{Connection between \RSLN{} and Geometric Phase} \label{sec:Linking_Phase}

In the following we will show that the acquired Berry phase along the momentum space circle can be understood as a linking of the phasor trajectory with the momentum space circle.
To this end we go back to the general Fourier transform of the vector wave field $\tilde{\mathbf{V}}(k,\chi)$, where, as in the main manuscript, $k$ and $\chi$ are the cylindrical coordinates in momentum space.
As discussed in the main manuscript, for monochromatic waves, the intensity distribution of this Fourier transform is localized on a ring with $k=k_0$. We can express the normalized vector field $\tilde{\mathbf{V}}(\chi)/\tilde{|\mathbf{V}}(\chi)| =  \ \tilde{\mathbf{V}}_\mathrm{pol}(\chi) \exp\left(\mathrm{i} \ \phi(\chi)\right)$ only as a function of the angle in momentum space $\chi$.
The ring $C_1$ is the first of the two considered linked curves; the base curve $\mathbf{c}_\mathrm{1}(\chi)= k_0 \,\hat{\mathbf{k}}(\chi)$.
As we consider a spin momentum locked wave field, the polarization state in momentum space is only related to the reciprocal angle $\chi$. Therefore, the topology of the vector field in momentum space is carried by the phasor $\exp(\mathrm{i} \,\phi(\chi))$. The phase $\phi(\chi)$ can be used to construct the second curve $C_2$ with $\mathbf{c}_\mathrm{2}(\chi) = \mathbf{c}_\mathrm{1}(\chi)+ k_0\left(\sin\phi(\chi)\,\hat{\mathbf{k}}+\cos\phi(\chi)\,\hat{\mathbf{k}}_{z}\right)$ where $\hat{\mathbf{k}}_{z}$ is the $z$-unit vector in cylindrical coordinates in momentum space.
Such a linking of two oriented curves $C_1$ and $ C_2$ is generally described by the Gauss linking number; this link then carries the topology of the phasor.
The Gauss linking number between the base curve $C_\mathrm{1}$ and the phasor curve $C_\mathrm{2}$ can be calculated using
\begin{equation}
    \mathrm{link}(C_1,C_2) = \oint_{C_1}\oint_{C_2}\mathrm{d}\chi\,\mathrm{d}\chi'\,\frac{\mathbf{c}_\mathrm{1}(\chi)-\mathbf{c}_\mathrm{2}(\chi')}{|\mathbf{c}_\mathrm{1}(\chi)-\mathbf{c}_\mathrm{2}(\chi')|^3}\cdot\left(\partial_{\chi}{\mathbf{c}}_\mathrm{1}(\chi)\times \partial_{\chi'}{\mathbf{c}}_\mathrm{2}(\chi')\right). \label{eq:LinkingIntegral}
\end{equation}

The nature of this linking is schematically depicted in Fig.~\ref{fig:Figure_S3}A, with the base curve $C_1$ in red and $C_2$ in blue, respectively. A geometric argument is sufficient to demonstrate that the winding number associated with the Berry phase is equal to the momentum space linking number (\RSLN{}): We smoothly deform one of the curves into a straight line so that the other curve winds around this straight line. This situation is depicted in Fig.~\ref{fig:Figure_S3}B. Note that both curves are still closed, as indicated by the dashed lines.
In this picture, the linking number of the red and blue curves is given by the number of times the blue curve winds around the central red curve. 
Thus, from this geometrical consideration, we have demonstrated the equality of the winding number and the linking number $\mathrm{link}(C_1,C_2)$ for the phasor trajectory in momentum space. Therefore, the \RSLN{} is equal to the winding number $\mathrm{link}(C_1,C_2)=\mathcal{N}$.

\subsubsection*{Connection between \RSLN{} and OAM} \label{sec:OAM_Linking}

Figure 2 of the main manuscript discusses the special case of pure cylindrical harmonics. For these fields, the OAM expectation value $\mathcal{J}_z$ for integer excitation phase steps is equal to the \RSLN{}.
Here, we will derive analytically that for OAM modes with a single isolated phase vortex in position space, the expression for the OAM expectation value in momentum space recovers the expression for the \RSLN{} up to a factor of $\hbar$.
The OAM expectation value \cite{Belinfante1940, Davis2023} of the here considered spin-momentum locked wave field is given in cylindrical coordinates with angle $\varphi$ as 
\begin{equation}
    \mathcal{J}_z = \frac{\int \mathrm{d}^2r \, V_z^{*}(\mathbf{r})(-\mathrm{i}\hbar\,\partial_\varphi) V_z(\mathbf{r})}{\int\mathrm{d}^2r \, V_z^{*}(\mathbf{r}) V_z(\mathbf{r})} \label{eq:OAMinRealspace}.
\end{equation}
It should be noted that for the here considered spin-momentum locked fields, the out of plane-component $V_z$ is used to calculate the OAM.
We will transform this position space based formulation into a momentum space based formulation.
For readability purposes, we will only consider the numerator $\mathcal{J}^{\mathrm{num}}_z$ here; the denominator is present for intensity normalization and therefore transforms straightforward into momentum space.
We express the field $V_z(\mathbf{r)}$ via the inverse Fourier transformation of its momentum space representation $\tilde{V}_z(k_x,k_y)$ and perform the angular derivative $\partial_\varphi=x\partial_y-y\partial_x$ in Cartesian coordinates, yielding
\begin{equation}
    \mathcal{J}^{\mathrm{num}}_z=\frac{\mathrm{-\hbar}}{4\pi^2}\int\mathrm{d}^2r\int\mathrm{d}^2k\int\mathrm{d}^2k' \left(x k_y'-y k_x'\right)\tilde{V}^{*}_z(k_x,k_y)\tilde{V}_z(k'_x,k'_y)\,\mathrm{e}^{\mathrm{i} \,\mathbf{r}\cdot\left(\mathbf{k}-\mathbf{k'}\right)}.
\end{equation}
To obtain a formulation of the OAM expectation value in momentum space, we first evaluate the integral over position space. Only then, one of the integrals over momentum space can collapse, yielding a formulation of the OAM expectation value directly in momentum space.
To this extend, we use that $x\,\mathrm{exp}(ax)=1/a\,\partial_a\mathrm{exp}(ax)$, allowing us to first rewrite the integral expression and then calculate it explicitly, thus obtaining
\begin{equation}
    \begin{split}
        \mathcal{J}^{\mathrm{num}}_z &= \frac{\mathrm{-\hbar}}{4\pi^2}\int\mathrm{d}^2r\,\int\mathrm{d}^2k\,\int\mathrm{d}^2k'\,\tilde{V}^{*}_z(k_x,k_y)\tilde{V}_z(k'_x,k'_y) \left(-\mathrm{i}\,\partial_{k_x} k_y'+\mathrm{i}\,\partial_{k_y}  k_x'\right)\mathrm{e}^{i\,\mathbf{r}\cdot\left(\mathbf{k}-\mathbf{k'}\right)} \\
        & = \mathrm{-i\hbar}\int\mathrm{d}^2k\,\int\mathrm{d}^2k'\,\tilde{V}^{*}_z(k_x,k_y)\tilde{V}_z(k'_x,k'_y) \left(k_x'\partial_{k_y}  -k_y'\partial_{k_x} \right)\delta(\mathbf{k}-\mathbf{k'}) 
    \end{split}
\end{equation}
The integral over the momentum coordinate $\mathbf{k'}$can be evaluated using the integration by parts identities for $\partial_{k_i}\delta(k_i-k_i')$.
\begin{equation}
    \begin{split}
        \mathcal{J}^{\mathrm{num}}_z 
        & = \mathrm{-i\hbar}\int\mathrm{d}^2k\,\int\mathrm{d}^2k'\,\tilde{V}^{*}_z(k_x,k_y)\tilde{V}_z(k'_x,k'_y) \left(k_x'\partial_{k_y}  -k_y'\partial_{k_x} \right)\delta(\mathbf{k}-\mathbf{k'}) \\
                & = \mathrm{-i\hbar}\int\mathrm{d}^2k\,\tilde{V}^{*}_z(k_x,k_y) \left(k_x\partial_{k_y}  -k_y\partial_{k_x} \right) \tilde{V}_z(k_x,k_y)
        \end{split}
\end{equation}
Similar to the angular derivative in position space we can identify the angular derivative in momentum space with its cartesian representation $\partial_\chi=k_x\partial_{k_y}-k_y\partial_{k_x}$. We thus arrive at
\begin{equation}
    \begin{split}
        \mathcal{J}_z^\mathrm{num} &= \mathrm{-i\hbar}\int\mathrm{d}^2k\,\tilde{V}^{*}_z(k,\chi) \partial_\chi\tilde{V}_z(k,\chi)\\
    &= \hbar\, \int\mathrm{d}^2k \,\mathrm{Im}\left[\tilde{V}^{*}_z(k,\chi)\partial_\chi\tilde{V}_z(k,\chi)\right]-\frac{\mathrm{i}}{2}\partial_\chi\left|\tilde{V}_z(k,\chi)\right|^2.
    \end{split}
\end{equation}
Here we expressed the integrand via its real and imaginary parts
\begin{equation}
    \begin{split}
    -\mathrm{i} \, \tilde{V}^{*}_z(k,\chi) \partial_\chi\tilde{V}_z(k,\chi)&=\mathrm{Im}\left[\tilde{V}^{*}_z(k,\chi) \partial_\chi\tilde{V}_z(k,\chi)\right]-\mathrm{i}\,\mathrm{Re}\left[\tilde{V}^{*}_z(k,\chi) \partial_\chi\tilde{V}_z(k,\chi)\right]\\
    &=\mathrm{Im}\left[\tilde{V}^{*}_z(k,\chi)\partial_\chi\tilde{V}_z(k,\chi)\right]-\frac{\mathrm{i}}{2}\partial_\chi\left|\tilde{V}_z(k,\chi)\right|^2.
        \end{split}
\end{equation}
As $|\tilde{V}_z(k,\chi)|^2$ is periodic in $\chi$, the integral over its angular derivative must vanish. Therefore, we arrive at
\begin{equation}
    \begin{split}
        \mathcal{J}_z^\mathrm{num} 
    &= \hbar\,\mathrm{Im}\left[ \int\mathrm{d}^2k \,\tilde{V}^{*}_z(k,\chi)\partial_\chi\tilde{V}_z(k,\chi)\right].
    \end{split}
\end{equation}
Combining this result with the intensity normalization in the denominator, we obtain the expression for the OAM expectation value in momentum space given by
\begin{equation}
    \begin{split}
        \mathcal{J}_z &=\hbar \, \mathrm{Im}\left[\frac{\int\mathrm{d}^2k\,\tilde{V}^{*}_z(k,\chi)\partial_\chi \tilde{V}_z(k,\chi)}{\int\mathrm{d}^2k \, \tilde{V}^{*}_z(k,\chi) \tilde{V}_z(k,\chi)}\right].
    \end{split}
\end{equation}
For the monochromatic wave field considered in the main manuscript, the momentum space intensity is localized on a circle in momentum space, i. e., $\tilde{V}_z\propto\delta(k-k_{0})$. This effective reduction of dimensionality further reduces the expression for the OAM to a single integral over the reciprocal angle $\chi$ 
\begin{equation}
   \mathcal{J}_z = \hbar\, \mathrm{Im}\left[\frac{\int_0^{2\pi}\mathrm{d}\chi \tilde{V}^{*}_z(k_\mathrm{0},\chi)\partial_\chi\tilde{V}_z(k_\mathrm{0},\chi)}{\int_0^{2\pi}\mathrm{d}\chi \, \tilde{V}^{*}_z(k_0,\chi) \tilde{V}_z(k_0,\chi)}\right].
\end{equation}
The complex field component $\tilde{V}_z(k_\mathrm{0},\chi)=\tilde{V}_{\mathrm{amp},z}(k_\mathrm{0},\chi)\exp{(\mathrm{i} \, \phi(\chi))}$ can be represented in polar form  with real amplitude $\tilde{V}_{\mathrm{amp},z}$ and the phasor $\exp{(\mathrm{i} \, \phi(\chi))}$  with the phase $\phi(\chi)\in\mathbb{R}$.
Inserting this polar representation into the OAM expectation value yields
\begin{equation}
    \mathcal{J}_z = \hbar\,\mathrm{Im}\left[\frac{\mathrm{i}\int_0^{2\pi}\mathrm{d}\chi \,|\tilde{V}_{\mathrm{amp},z}(k_\mathrm{0},\chi)|^2\partial_\chi\phi(\chi)}{\int_0^{2\pi}\mathrm{d}\chi \,|\tilde{V}_{\mathrm{amp},z}(k_0,\chi)|^2}+\frac{\int_0^{2\pi}\mathrm{d}\chi\,\tilde{V}_{\mathrm{amp},z}(k_\mathrm{0},\chi)\partial_{\chi}\tilde{V}_{\mathrm{amp},z}(k_\mathrm{0},\chi)}{\int_0^{2\pi}\mathrm{d}\chi \, |\tilde{V}_{\mathrm{amp},z}(k_0,\chi)|^2}\right].
\end{equation}
The second term is purely real and, therefore, does not contribute to the OAM expectation value. The first term is purely imaginary and represents the intensity weighted phase change along the momentum space circle. It is important to note here that it is the intensity weight of the phase gradient that breaks the general connection between OAM and \RSLN{} as discussed in the main manuscript.

In the special case of circular harmonic fields carrying integer OAM values with an isolated phase vortex in position space as discussed in the main manuscript, the amplitude along the momentum space circle does not change, i. e. $V_{\mathrm{amp},z}(k_\mathrm{0},\chi)=V_{\mathrm{amp},z}(k_\mathrm{0})$. Therefore, we can separate the integral over the derivative of the phase from the intensity and obtain 
\begin{equation}
    \mathcal{J}_z = \frac{\hbar}{2\pi}\int_0^{2\pi}\mathrm{d}\chi\,\partial_\chi\phi(\chi),
\end{equation}
which corresponds to the expression for the Berry phase given in Eq.~\ref{eq:gemoetric_phase} up to a factor of $\hbar/2\pi$. This relates the OAM expectation to the \RSLN{}, that is $\mathcal{N}=\mathcal{J}_z/\hbar$.

\subsubsection*{Deformations of the Boundary}

The \RSLN{} remains invariant under deformations of the excitation boundary, as discussed in the context of figure~\ref{fig:figure_3} of the main manuscript. This section of the supplementary text analytically shows the invariance of the \RSLN{} under such deformations. 
The position space vector field on a two-dimensional plane originating from a source placed at $\mathbf{r}_\mathrm{b}$ can be described by a Hankel wavelet
\begin{equation}
     \mathbf{V}(\mathbf{r}, \mathbf{r}_\mathrm{b}) =\mathbf{V}_\mathrm{pol}(\mathbf{r}-\mathbf{r}_\mathrm{b})A(\mathbf{r}_\mathrm{b})H_0^{(1)}\left(k_\mathrm{0} \left|\mathbf{r}-\mathbf{r}_\mathrm{b}\right|\right).
 \end{equation}
Within the Hankel wavelet, $\mathbf{V}_\mathrm{pol}(\mathbf{r}-\mathbf{r}_\mathrm{b})$ is the polarization vector of the vector field at position~$\mathbf{r}$, $A(\mathbf{r}_\mathrm{b})$ is the amplitude of the field excited at $\mathbf{r}_\mathrm{b}$ and $H_0^{(1)}$ is the Hankel function of the first kind zeroth order. By superposing the fields excited at all source points along the boundary $\mathbf{r}_\mathrm{b}(\theta)$, parametrized by the angle $\theta$, we can obtain the overall position space vector field $\mathbf{V}(\mathbf{r})$.
\begin{equation}
     \mathbf{V}(\mathbf{r}) = \int\mathrm{d}\theta \int\mathrm{d}^2 r' \mathbf{V}_\mathrm{pol}(\mathbf{r}-\mathbf{r'})A(\mathbf{r}_\mathrm{b})\delta\left(\mathbf{r}'-\mathbf{r}_\mathrm{b}(\theta)\right)H_0^{(1)}\left(k_\mathrm{0} \left|\mathbf{r}-\mathbf{r}'\right|\right)|\dot{\mathbf{r}}_\mathrm{b}(\theta)|.
 \end{equation}
%We consider excitation structures that can be described as $\mathbf{r}_\mathrm{b}(\theta)=\left(R+\Delta R(\theta)\right)\hat{\mathbf{r}}(\theta)$,  where $R$ is the radius of a circular boundary and $\Delta R(\theta)$ describes any deviations from the circular boundary.%
Where, $|\dot{\mathbf{r}}_\mathrm{b}(\theta)|$ is the length element of the boundary parametrized by $\theta$, with the dot denoting the derivative with respect to $\theta$.
We now perform the integral over $r'$ and spatially Fourier transform $\mathbf{V}(\mathbf{r})$, arriving at
\begin{equation}
    \tilde{\mathbf{V}}(\mathbf{k}) = \int \mathrm{d}\theta\,\int\mathrm{d}^2 r\, \mathrm{e}^{\mathrm{i} \,\mathbf{k}\cdot\mathbf{r}} \,\mathbf{V}_\mathrm{pol}(\mathbf{r}-\mathbf{r}_\mathrm{b})A(\mathbf{r}_\mathrm{b})H_0^{(1)}\left(k_\mathrm{0} \left|\mathbf{r}-\mathbf{r}_\mathrm{b}(\theta)\right|\right)|\dot{\mathbf{r}}_\mathrm{b}(\theta)|. \label{eq:BoundaryDeformFTStart}
\end{equation}
We consider excitation structures that can be described by $\mathbf{r}_\mathrm{b}(\theta)=\left(R+\Delta R(\theta)\right)\hat{\mathbf{r}}(\theta)$, and assuming an excitation structure much larger than the considered region within the boundary $r\ll r_\mathrm{b}$, we find $|\mathbf{r}-\mathbf{r}_\mathrm{b}| \approx R+\Delta R-r\cos{\left(\theta-\varphi\right)}$, where the position $\mathbf{r}$ is given in polar coordinates, with radius $r$ and angle $\varphi$.
To further simplify the expression above, we can approximate $H_0^{(1)}\left(k_\mathrm{0} \left|\mathbf{r}-\mathbf{r}_\mathrm{b}\right|\right)$ far away from the excitation source as:
\begin{equation}
    H_0^{(1)}\left(k_\mathrm{0} \left|\mathbf{r}-\mathbf{r}_\mathrm{b}\right|\right) \approx \sqrt{\frac{2}{k_\mathrm{0} \pi}} \sqrt{\frac{1}{R+\Delta R(\theta)}}\mathrm{e}^{\mathrm{i} \,k_\mathrm{0}\left(R+\Delta R-r\cos{\left(\theta-\varphi\right)}\right)}.
\end{equation}
We express $\mathbf{k}$ in polar coordinates, with $k$ as the distance from the origin in momentum space and $\chi$  as the angle in momentum space. Using $\mathbf{k}\cdot\mathbf{r}=k\, r \cos{\left(\chi-\varphi\right)}$ and $|\dot{\mathbf{r}}_\mathrm{b}(\theta)|=\sqrt{\left(R+\Delta R(\theta)\right)^2+\dot{\Delta R(\theta)}^2}$, equation \ref{eq:BoundaryDeformFTStart} can be written as
%the vector field in momentum space representation as
\begin{equation}
    \begin{split}
        \tilde{\mathbf{V}}(k,\chi) = \sqrt{\frac{2}{k_\mathrm{0}\pi}}&\int\mathrm{d}\theta\, \left[ \frac{\sqrt{\left(R+\Delta R(\theta)\right)^2+\dot{\Delta R(\theta)}^2}}{\sqrt{R+\Delta R(\theta)}} \mathrm{e}^{\mathrm{i} \,k_\mathrm{0}\left(R+\Delta R(\theta)\right)}\mathbf{V}_\mathrm{pol}(\theta)A(\theta) \right.\\
        &\left.\int_0^{\infty} \mathrm{d}r\, r\int_0^{2\pi}\mathrm{d}\varphi\, \mathrm{e}^{\mathrm{i} \,k r \cos{\left(\chi-\varphi\right)}}\mathrm{e}^{-\mathrm{i} \,k_\mathrm{0} r \cos{\left(\theta-\varphi\right)}} \right]\label{eq:fourierTransformElectricFieldStart}.
    \end{split}
\end{equation}
Note here that for $r\ll r_\mathrm{b}$ the polarization vector $\mathbf{V}_\mathrm{pol}(-\mathbf{r}_\mathrm{b})$ and the amplitude $A(\mathbf{r}_\mathrm{b})$ only depend on the boundary angle $\theta$, therefore $\mathbf{V}_\mathrm{pol}(-\mathbf{r}_\mathrm{b})=\mathbf{V}_\mathrm{pol}(\theta)$ and $A(\mathbf{r}_\mathrm{b})=A(\theta)$, respectively.
This expression separates into two parts: The first part only depends on the angle around the boundary $\theta$. The second part depends on both, the real space coordinates $r$, $\varphi$ , and the boundary angle $\theta$.
To obtain the momentum space representation of the vector field, we first calculate the integrals over real space and then proceed to the integral over $\theta$.
Using the Jacobi-Anger identity allows us to express the exponential functions as sums over Bessel functions of first kind:
\begin{equation}
    \begin{split}
        \mathrm{e}^{\mathrm{i} \,k r \cos{\left(\chi-\varphi\right)}} &=  \sum_{n=-\infty}^{\infty} \mathrm{i} ^n J_n(kr) \mathrm{e}^{\mathrm{i}n\left(\chi-\varphi\right)}, \\
        \mathrm{e}^{-\mathrm{i}\,k_\mathrm{0} r \cos{\left(\theta-\varphi\right)}} &=  \sum_{m=-\infty}^{\infty} (\mathrm{-i})^m J_m(k_\mathrm{0} r) \mathrm{e}^{\mathrm{i}m\left(\varphi-\theta\right)}.
    \end{split}
\end{equation}
With this, the second part $\mathrm{II}$ of Eq. (\ref{eq:fourierTransformElectricFieldStart}) expands to
\begin{equation}
    \begin{split}
\mathrm{II}  =& \int_0^\infty\mathrm{d}r\, r\int_0^{2\pi}\mathrm{d}\varphi \,\mathrm{e}^{\mathrm{i}k r \cos{\left(\chi-\varphi\right)}}\mathrm{e}^{-\mathrm{i}k_\mathrm{0} r \cos{\left(\theta-\varphi\right)}} \label{eq:fourierTransformElectricFieldReal}\\
        =& \int_0^\infty \mathrm{d}r \,r \int_0^{2\pi} \mathrm{d}\varphi \sum_{n,m} \mathrm{i}^{n}(-\mathrm{i})^m J_n(k r)J_m(k_\mathrm{0} r) \mathrm{e}^{\mathrm{i} n\left(\chi-\varphi\right)}\mathrm{e}^{\mathrm{i} m \left(\varphi-\theta\right)} \\
        =& \int_0^\infty \mathrm{d}r \,r \sum_{n,m} \mathrm{i}^{n}(-\mathrm{i})^m J_n(k r)J_m(k_\mathrm{0} r)\mathrm{e}^{\mathrm{i} n\chi}\mathrm{e}^{\mathrm{i} m\theta}\int_0^{2\pi}\mathrm{d}\varphi \,\mathrm{e}^{\mathrm{i}\varphi\left(m-n\right)}.
    \end{split}
\end{equation}
Evaluating the angular integral, we obtain
\begin{equation}
    \begin{split}
\mathrm{II}
    = 2\pi\int_0^{\infty}  \mathrm{d}r\, r\sum_{n,m} \mathrm{i}^{n}(-\mathrm{i})^m J_n\left(k r\right)J_m\left(k_\mathrm{0} r\right)\mathrm{e}^{\mathrm{i} \left(n\chi-m\theta\right)}\delta_{nm}.
    \end{split}
\end{equation}
Note that $\delta_{nm}$ is the Kronecker delta with $\delta_{nm}=0$ if $n\neq m$ and $\delta_{nn}=1$, collapsing the double sum over $n,m$ to a single sum over $n=m$.  
\begin{equation}
    \begin{split}
\mathrm{II}
        =& 2\pi\sum_{n} \mathrm{e}^{\mathrm{i} n\left(\chi-\theta\right)}\int_0^{\infty}  \mathrm{d}r \, r J_n\left(k r\right)J_n\left(k_\mathrm{0} r\right) \\
    \end{split}
\end{equation}
The orthogonality of Bessel-functions of the same order with respect to their argument  $\int _0^{\infty} \mathrm{d}r\, r J_{n}(k r)J_{n}(k_\mathrm{0} r)=1/k_\mathrm{0} \,\delta(k-k_\mathrm{0})$ allows us to evaluate the integral over $r$.
\begin{equation}
    \begin{split}
   \mathrm{II}=& \frac{2\pi}{k_\mathrm{0}}\sum_{n} \mathrm{e}^{\mathrm{i} n\left(\chi-\theta\right)}\delta(k-k_\mathrm{0}) \\
    \end{split}
\end{equation}
The sum of the exponential functions can be identified as a Dirac delta, therefore yielding
\begin{equation}
    \begin{split}
            \mathrm{II}
        =& \frac{4\pi^2}{k_\mathrm{0}}\delta(k-k_\mathrm{0})\delta(\chi-\theta).
    \end{split}
\end{equation}
Inserting this result into the momentum space representation of the vector field given in Eq. (\ref{eq:fourierTransformElectricFieldStart}), we can evaluate the integral over the boundary angle 
\begin{equation}
    \begin{split}
        \tilde{\mathbf{V}}(k,\chi) &= \sqrt{\frac{32\pi^3}{k_\mathrm{0}^3}} \int_0^{2\pi} \mathrm{d}\theta \frac{\sqrt{\left(R+\Delta R(\theta)\right)^2+\dot{\Delta R(\theta)}^2}}{\sqrt{R+\Delta R(\theta)}} \mathrm{e}^{\mathrm{i}k_\mathrm{0}\left(R+\Delta R(\theta)\right)}\mathbf{V}_\mathrm{pol}(\theta)A(\theta) \delta\left(\chi-\theta\right)\delta(k-k_\mathrm{0}) \\
        &= \sqrt{\frac{32\pi^3}{k_\mathrm{0}^3}}\frac{\sqrt{\left(R+\Delta R(\chi)\right)^2+\dot{\Delta R(\chi)}^2}}{\sqrt{R+\Delta R(\chi)}}\mathrm{e}^{\mathrm{i}k_\mathrm{0}\left(R+\Delta R(\chi)\right)}\tilde{\mathbf{V}}_\mathrm{pol}(\chi)A(\chi)\delta(k-k_\mathrm{0}) \label{eq:fourierTransformElectricFieldEnd}
    \end{split}
\end{equation}
This is the general expression for the Fourier transform of a monochromatic vector field in a deformed circular excitation structure. %As the boundary is assumed to be far away from the observed field, the angle dependence of the polarization vector in position and momentum space is equal, i.e. $\mathbf{V}_\mathrm{pol}(\chi)=\tilde{\mathbf{V}}_\mathrm{pol}(\chi)$.%
Note, that the Fourier-transformed field is entirely localized on a ring at $k=k_\mathrm{0}$. 
Irrespective of the excitation boundary, forms a compact ring ($S^1$) in momentum space.
Connecting back to the previously used momentum space phasor representation of the field $\tilde{\mathbf V}(k,\chi)= \tilde{\mathbf V}_\mathrm{amp}(k,\chi)\exp{(\mathrm{i}\, \phi(\chi)})$ we can identify the amplitude vector and phasor:
\begin{equation}
    \begin{split}
    \tilde{\mathbf V}_\mathrm{amp}(k,\chi) &= \sqrt{\frac{32\pi^3}{k_\mathrm{0}^3}}\frac{\sqrt{\left(R+\Delta R(\chi)\right)^2+\dot{\Delta R(\chi)}^2}}{\sqrt{R+\Delta R(\chi)}}\tilde{\mathbf{V}}_\mathrm{pol}(\chi)A(\chi)\delta(k-k_\mathrm{0}) \\
    e^{\mathrm{i} \phi(\chi)} &= \mathrm{e}^{\mathrm{i}k_\mathrm{0}\left(R+\Delta R(\chi)\right)} \\
    \end{split}
\end{equation}
Calculating the \RSLN{} from the phasor, one obtains
\begin{equation}
    \mathcal{N} = k_\mathrm{
    0
    }\oint \mathrm{d}\chi\, \partial_\chi\Delta R(\chi)
\end{equation}
Thus, a perfectly circular excitation boundary does not contribute to the accumulated geometric phase.
Further, we find that only discontinuous deformations can contribute to the \RSLN{}. Therefore, only boundaries containing a discontinuous deformation, such as spirals, can exhibit a non-trivial \RSLN{}.
Any deformation can be decomposed into $\Delta R(\chi)=S(\chi)+P(\chi)$, where $S(\chi)$ is the spiral-shaped part that carries the topology of the system and $P(\chi)$ is the periodic perturbation to the ideal excitation structure not contributing to the topology.
We have thus shown analytically that the \RSLN{} remains constant under continuous deformations of the excitation structure.

\newpage
%%%%%%%%%%%%%%%% SUPPLEMENTARY FIGURES %%%%%%%%%%%%%%%

\begin{figure}
    \centering
    \includegraphics[width=0.5\linewidth]{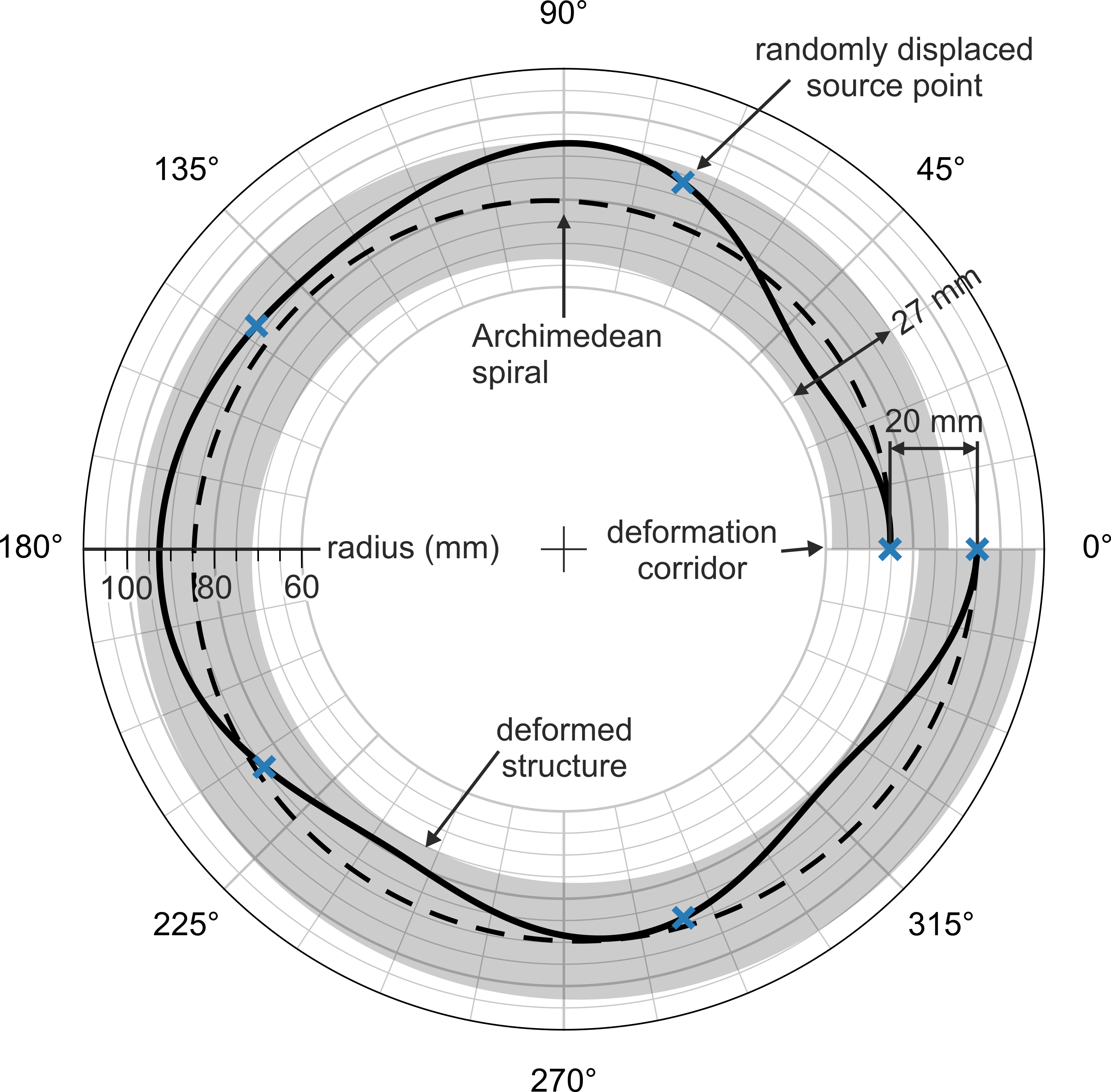}
    \caption{\textbf{Construction of the deformed structure.} The reference geometry is an Archimedean spiral (dashed line) with a 20 mm gap. Six source points (blue crosses) are sampled along the spiral. The first and last points remain fixed, while the remaining four are randomly displaced in the radial direction. A spline fitted through these points defines the deformed structure (solid line), which remains within a 27 mm wide deformation corridor (shaded region) around the reference spiral. Polar angle in degrees and radius in millimeters are indicated. }
    \label{fig:Figure_S2}
\end{figure}

\begin{figure}
    \centering
    \includegraphics[width=1\linewidth]{figures/Supplementary_Linking.png}
    \caption{\textbf{Geometrical construction of the equivalence between linking and winding number.} (\textbf{A}) Two oriented curves $C_1$ (in red) and $C_2$ (in blue) are linked once. (\textbf{B}) The red curve is fully  stretched out with the blue curve winding around it as a function of reciprocal angle $\chi$, the dashed lines indicate the closure of the two curves.}
    \label{fig:Figure_S3}
\end{figure} 

%%%%%%%%%%% CAPTIONS FOR OTHER SUPPLEMENTARY FILES %%%%%%%%%%

\clearpage % Clear all remaining figures and tables then start a new page

\paragraph{Caption for Movie S1.}
\textbf{Phasor trajectory as a function of excitation phase step.}
The phasor trajectory has been obtained from SPP modes carrying orbital angular momentum using NSOM. The movie is an animated version of the data points displayed in Fig.~\ref{fig:figure_2}H.

\paragraph{Caption for Movie S2.}
\textbf{Raw data of the deformed Archimedean spiral.}
The movie shows the raw data obtained by the water wave experiment in slow motion (1/8). The excited wave pattern carries an excitation phase step of $g=1$. The distortions of the checkerboard below the water surface allow for the reconstruction of the water height.

\paragraph{Caption for Movie S3.}
\textbf{Reconstructed displacement vectors of the water wave excited at the deformed Archimedean spiral.}
The movie shows the reconstructed displacement field vectors of the water surface depicted in Fig.~\ref{fig:figure_3}C.

\paragraph{Caption for Movie S4.}
\textbf{Phasor trajectory of the water wave excited at the deformed Archimedean spiral as a function of time.}
The movie shows the phasor trajectory of the data shown in Fig.~\ref{fig:figure_3}E as a function of time. The data shown for\RSLN{} $\mathcal{N}=1$ in Fig.~\ref{fig:figure_3}G is extracted from the phasor trajectory shown in this movie.

\end{document}